\documentclass[twocolumn,prb,preprintnumbers,amsmath,amssymb,superscriptaddress]{revtex4-1}
\usepackage[dvipdfmx]{graphicx}
\usepackage{mediabb}
\usepackage{color}
\usepackage{dcolumn}
\usepackage{bm}
\usepackage{amsmath,amssymb}
\usepackage{ulem}

\UseRawInputEncoding

\begin{document}

\title{
Spin excitations coupled with lattice and charge dynamics in La$_{2-x}$Sr$_{x}$CuO$_4$
}

\author{K. Ikeuchi}
 \email{k\_ikeuchi@cross.or.jp}
\affiliation{
Neutron Science and Technology Center,
Comprehensive Research Organization for Science and Society (CROSS), Tokai, Ibaraki 319-1106, Japan
}

\author{S. Wakimoto}
\affiliation{
Materials Sciences Research Center, Japan Atomic Energy Agency (JAEA),
Tokai, Ibaraki 319-1195, Japan
}

\author{M. Fujita}
\affiliation{
Institute for Materials Research, Tohoku University,
Katahira, Sendai 980-8577, Japan
}

\author{T. Fukuda}
\affiliation{
Material Science Research Center, Japan Atomic Energy Agency (JAEA), Ko-to, Hyogo 679-5148, Japan
}

\author{R. Kajimoto}
\affiliation{
Materials and Life Science Division (MLF),
J-PARC Center, Tokai, Ibaraki 319-1195, Japan
}

\author{M. Arai}
\affiliation{
Materials and Life Science Division (MLF),
J-PARC Center, Tokai, Ibaraki 319-1195, Japan
}
\affiliation{
European Spallation Source ERIC, P.O. Box 176, SE-221 00 Lund, Sweden
}

\date{\today}

\begin{abstract}

Spin excitations of layered copper oxide show various characteristic features,  depending on the carrier concentration. In this study, we conducted inelastic neutron scattering (INS) measurements on La$_{2-x}$Sr$_{x}$CuO$_4$ (LSCO), with $x$ = 0, 0.075, 0.18, and 0.30 and La$_{2-x}$Sr$_{x}$NiO$_4$ (LSNO) with 1/3, to clarify the origin of the intensity enhancement in the excitation spectrum of LSCO at the energy ($\omega$) of 16--19 meV [Phys. Rev. B {\bf 91}, 224404 (2015), {\it ibid}. {\bf 93}, 094416 (2016)]. 
We confirmed the presence of a peak-structure in the $\omega$-dependence of the local spin susceptibility $\chi''(\omega)$ of superconducting (SC) LSCO with a peak energy of 16--19 meV, where the spin excitations intersect optical phonon branches. A comparable peak-structure is not observed in the insulating La$_2$CuO$_4$, LSNO, and heavily overdoped LSCO with $x$ = 0.30. A dome-shaped $x$-dependence of the integrated intensity around the peak energies is revealed for SC phase by summarizing the present and previously reported results. 
Furthermore, our phonon calculation on LCO shows the existence of two optical branches at $\sim$19 meV that could stabilize stripe-alignment of carriers due to out-of-plane vibrations of Cu or O of the CuO$_2$ planes. 
These results indicate the interplay among spin, charge, and lattice dynamics and suggest that the intensity enhancement is associated with their composite excitations. 
\end{abstract}

\maketitle

\section{Introduction}

The elucidation of spin excitations in superconducting (SC) cuprate oxides is an important research issue related to doped Mott insulators. 
Extensive inelastic neutron scattering (INS) experiments have revealed a variety of spin excitations in cuprate superconductors, depending on the carrier concentrations~\cite{fujita_rev}. The undoped antiferromagnet La$_2$CuO$_4$ (LCO) shows spin-wave dispersion, which is well understood in the framework of spin-wave theory~\cite{headings_lco}. This excitation is replaced by the "hourglass"-shaped excitations in SC La$_{2-x}$Sr$_{x}$CuO$_4$ (LSCO)~\cite{vignolle_HG}, originally observed in La$_{1.875}$Ba$_{0.125}$CuO$_4$~\cite{tranquada_lbco} and YBa$_2$Cu$_3$O$_7$ (YBCO)~\cite{Arai1999, mook_ybco}. 
The hourglass excitations consist of a vertically standing low-energy incommensurate (IC) component and a high-energy outwardly dispersive component in a wide energy ($\omega$) spanning the waist energy at approximately 40 meV~\cite{goka_x005_HG, vignolle_HG, matsuda_HG}. The low-energy IC excitations evolve in the underdoped (UD) region upon doping with the increase in incommensurability ($\delta$) and degrade coincidentally with the suppression of superconductivity in the overdoped (OD) region~\cite{wakimoto_OD1, wakimoto_OD2, lipscombe_OD}. 
These experimental facts indicate an close relationship between the IC spin excitations and superconductivity. Intriguingly, corresponding measurements obtained by neutron and x-ray~\cite{braicovich_rixs, wakimoto_rixs} beams clarified a weak doping-dependence of the high-energy dispersion in LSCO~\cite{dean_xdep} and YBCO~\cite{letacon} up to a heavily OD regime. 

Lipscombe et al.~\cite{lipscombe_localchi} clarified the two-component character of spin excitations for LSCO with $x$ = 0.075, which shows intensity maximums in the $\omega$-dependence of local spin susceptibility ($\chi ''(\omega)$) at approximately $\omega$ = 19 and 40 meV. 
Recently, Sato et al.~\cite{Sato2020} reported the possible co-existence of IC and commensurate (C) excitations around the waist energy of the hourglass excitations in UD $x$ = 0.10 and optimally-doped (OP) $x$ = 0.16. An analysis based on a two-component picture for the high-quality INS data revealed evidence of the itinerant electron spin nature and localized magnetism for the low-energy IC and high-energy C excitations, respectively. 

Considering the one-dimensional alignment of spin and hole domains within a stripe model, the intensity maximum at 40 meV is interpreted as the saddle point character of the gapped excitation from an even-leg spin ladders system~\cite{tranquada_lbco, Vojita2004}. 
For the excitations showing an intensity maximum at a lower $\omega$ (attributed as intensity enhancement), polarized INS measurements on La$_{1.84}$Sr$_{0.16}$Cu$_{0.96}$Ni$_{0.04}$O$_4$ confirmed that the enhanced spectral weight is magnetic in origin~\cite{Matsuura2012}.
More recently, Wagman et al. ~\cite{wagman_enhance} pointed out that the spin excitations at the peak-energy of $\sim$19 meV cross with optical phonon branches for La$_{2-x}$Ba$_{x}$CuO$_4$ (LBCO) with $x$ = 0.035. They argued that the intensity enhancement was associated with a modification of the exchange coupling constant $J$ using a vibration mode of Cu-O-Cu bond. A similar peak-structure was also reported for the UD region of LBCO and LSCO~\cite{Wagman2016}. 

Motivated studies presented above, the purpose of this study is to gain insight into the relationship between intensity enhancement and superconductivity relating to spin excitations and clarifying the necessary conditions for intensity enhancement. 
To accomplish this, we performed INS measurements on LSCO with UD $x$ = 0.075 (LSCO(7.5)) and slightly OD $x$ = 0.18 (LSCO(18)), as well as non-SC $x$ = 0.30 (LSCO(30)). LSCO(30) shows no clear magnetic signal up to 40 meV~\cite{wakimoto_OD2}, thus we measured the phonons in this compound as the reference. The presence or absence of intensity enhancement was also examined for insulating systems of undoped antiferromagnet LCO and stripe-ordered La$_{2-x}$Sr$_{x}$NiO$_4$ with $x$ = 1/3 (LSNO(1/3)).

We confirmed the enhancement of the magnetic spectral weight in LSCO(7.5) at 16 meV and LSCO(18) at 19 meV, where optical phonon branches are lying. Such enhancement is absent in the insulating LCO and LSNO(1/3), even though the same phonon branches exist. The energy-integrated $\chi ''(\omega)$ between the range of 12 and 28 meV covering the peak energies ($E_{\rm p}$) exhibits a dome-shaped $x$-dependence in the SC phase. To assign the optical phonons at approximately 19 meV that could couple with charge stripes, we performed a phonon calculation on LCO. Two possible modes containing out-of-plane oxygen vibration were clarified. 
These results suggest that the intensity enhancement is prone to appear through the interplay among spin, charge, and lattice dynamics in cuprates. 

The remainder of this paper is organized as follows. Section II outlines the INS measurements, sample preparation, and phonon calculations. The results are presented in Section III. A discussion of the possible origin of the intensity enhancement is presented in Section IV, and a summary is provided in Section V.

\section{Experiments and calculations}
Single crystals of LSCO and LSNO were grown using the traveling-solvent floating-zone method. The grown single crystals were 8 and 6 mm in diameter for LSCO and LSNO, respectively, with a length of 40 mm. Five to eight crystals used for the INS experiments were co-aligned for each compound by either a backscattering Laue method using an x-ray or a transparent Laue method using a $\gamma$-ray.

The INS experiments were performed using 4SEASONS, a time-of-flight (TOF) Fermi chopper spectrometer, installed at the Materials and Life Science Experimental Facility (MLF) of J-PARC~\cite{kajimoto, inamura}. Co-aligned samples were mounted on a closed-cycle refrigerator so that the $a$- and $c$-axes of the orthorhombic crystallographic notation were placed horizontally. The monochromatizing chopper (Fermi chopper) was tuned to obtain the neutron incident energies of $E_i$ = 48 and 55.47 meV for LSCO and LSNO(1/3), respectively. The Fermi chopper's rotation frequency was set to 250 Hz, producing an energy resolution of $\sim$6 \% of each incident energy at the elastic position. 

Most of the previous INS measurements of spin excitations in LSCO using the TOF technique were done with a configuration where the $c$-axis was parallel to the incident neutrons, $k_i\,||\,c$ ($k_i$ : wave vector of incident neutrons)~\cite{fujita_rev, Yamada1995, Birgeneau2006}.
This configuration effectively measures the spin excitations of LSCO in the energy and momentum (${\bm Q}$ = ($H$, $K$, $L$)) spaces due to the low dimensionality of the spin correlations.
In this study, neutron cross-sections were measured at this condition for LSCO(18). Additionally, the scattering intensities for LCO, LSCO(7.5), and LSCO(30) and LSCO(1/3) were mapped out in the four-dimensional ${\bm Q}$-$\omega$ space. For the latter case, the samples were rotated in the horizontal $a$-$c$ plane from $-30$ to $+60$ degrees with respect to the position of $k_i\,||\,c$. Data were collected at 2.5-degree steps and for 20 minutes at each angle. We measured the neutron scattering intensity at 5 K for all samples and above 200 K for LCO, LSCO(7.5), and LSCO(30). Hereafter, we present our data in the orthorhombic notation where the inplane antiferromagnetic zone center in LCO ($\pi$, $\pi$) corresponds to ($H$, $K$) = (1, 0). We use the energy unit with $\hbar$ = 1 throughout the paper.

INS provides the dynamical susceptibility as a function of the three-dimensional momentum ${\bm Q}$ and $\omega$, $\chi ''({\bm Q}, \omega)$, by measuring the double differential cross-section defined as follows: 
\begin{eqnarray*}
{\displaystyle\frac{d^2\sigma}{d\Omega\,d\omega}} = {\displaystyle\frac{2(\gamma r_e)^2}{\pi g^2\mu_B^2}}{\displaystyle\frac{k_f}{k_i}}\left|F({\bm Q})\right|^2{\displaystyle\frac{\chi''({\bm Q}, \omega)}{1-{\rm exp}(-\beta\omega)}}
\end{eqnarray*}
where $(\gamma r_{\rm e})^2 = 0.2905$ barn\, sr$^{-1}$, $(k_f/k_i )$ is the ratio of the final and incident neutron wave vector, $|F({\bm Q})|^2$ is the anisotropic magnetic form factor for a Cu$^{2+}$ $d_{x^2-y^2}$ orbital, and $1/(1-{\rm exp}(-\beta\omega))$ is the Bose population factor. We converted the magnetic intensity of LSCO(7.5), LSCO(18), and LSNO(1/3) in the absolute unit by analyzing the incoherent elastic scattering intensity. For LCO and LSCO(30), the absolute intensity was estimated from the ratio of the phonon intensity at the zone boundary (1, 0.5, 3) with that of LSCO(7.5). The ratio among LCO, LSCO(7.5), and LSCO(30) was 0.33 : 1 : 0.48. 

To assign phonon modes near (1, 0), we calculated phonon dispersions for the parent LCO based on the shell model using the open source package OpenPhonon~\cite{openphonon, phonon_param}.

\begin{figure*}[t]
\begin{center}
	\includegraphics[scale=0.125]{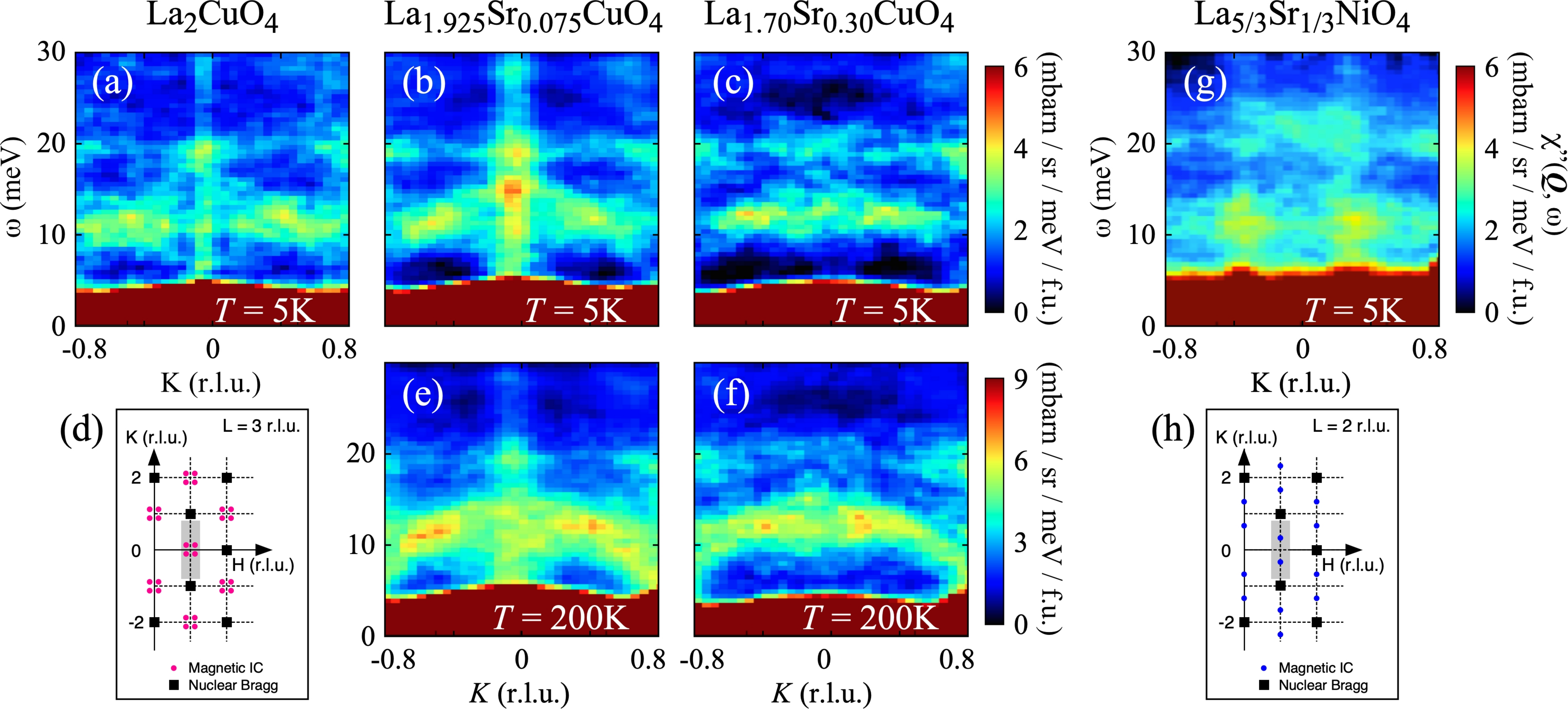}
\caption{
(Color online) Dynamical susceptibility in the $\omega$-$K$ plane for La$_{2-x}$Sr$_x$CuO$_4$ with: (a) $x$ = 0, (b) $x$ = 0.075, (c) $x$ = 0.30 at $T$ = 5 K, (e) $x$ = 0.075, (f) $x$ = 0.30 at $T$ = 200 K, and (g) La$_{2-x}$Sr$_x$NiO$_4$ with 1/3 at $T$ = 5 K.  The intensities were obtained by integrating over the ranges of $0.8 < H < 1.2$ and $2.5 < L < 3.5$.
The nuclear Bragg points and incommensurate magnetic positions in the $H$-$K$ plane for La$_{2-x}$Sr$_x$CuO$_4$ and La$_{2-x}$Sr$_x$NiO$_4$ are illustrated in panels (d) and (h), respectively. 
Shaded areas in (d) and (h) correspond to the integrated area in $K$ and range in $H$ of the horizontal axis for each intensity map. 
}
\label{fig1}
\end{center}
\end{figure*}

\section{Results}
Figure \ref{fig1} presents the overall neutron scattering intensity in the absolute units measured at 5 K for: (a) LCO, (b) LSCO(7.5), (c) LSCO(30), and (g) LSNO(1/3). Intensity maps at 200 K are also shown for (e) LSCO(7.5) and (f) LSCO(30). 
Data were integrated over the ranges of $0.8 < H < 1.2$ and $2.5 < L < 3.5$ so that all magnetic intensities around (1, 0) were included, as shown in Figs. \ref{fig1}(d) and (h). Low-energy spin excitations vertically stand along the energy direction at $K$ = 0 in LCO and LSCO(7.5), while similar magnetic signals were not observed in LSCO(30). Due to the insufficient instrumental resolution to resolve IC structure with a small value of $\delta$, a broad single peak centered at $K$ = 0 was detected for the LSCO(7.5) sample. In contrast, LSNO(1/3) shows well-defined IC signals due to the large $\delta$. More importantly, at a low temperature, relatively large intensities were observed for LSCO(7.5) at $K$ = 0 and $\omega$ = 16--19 meV, where optical phonon branches cross the spin excitations. This intensity enhancement disappears at 200 K. 

\begin{figure}[t]
\includegraphics[scale=0.12]{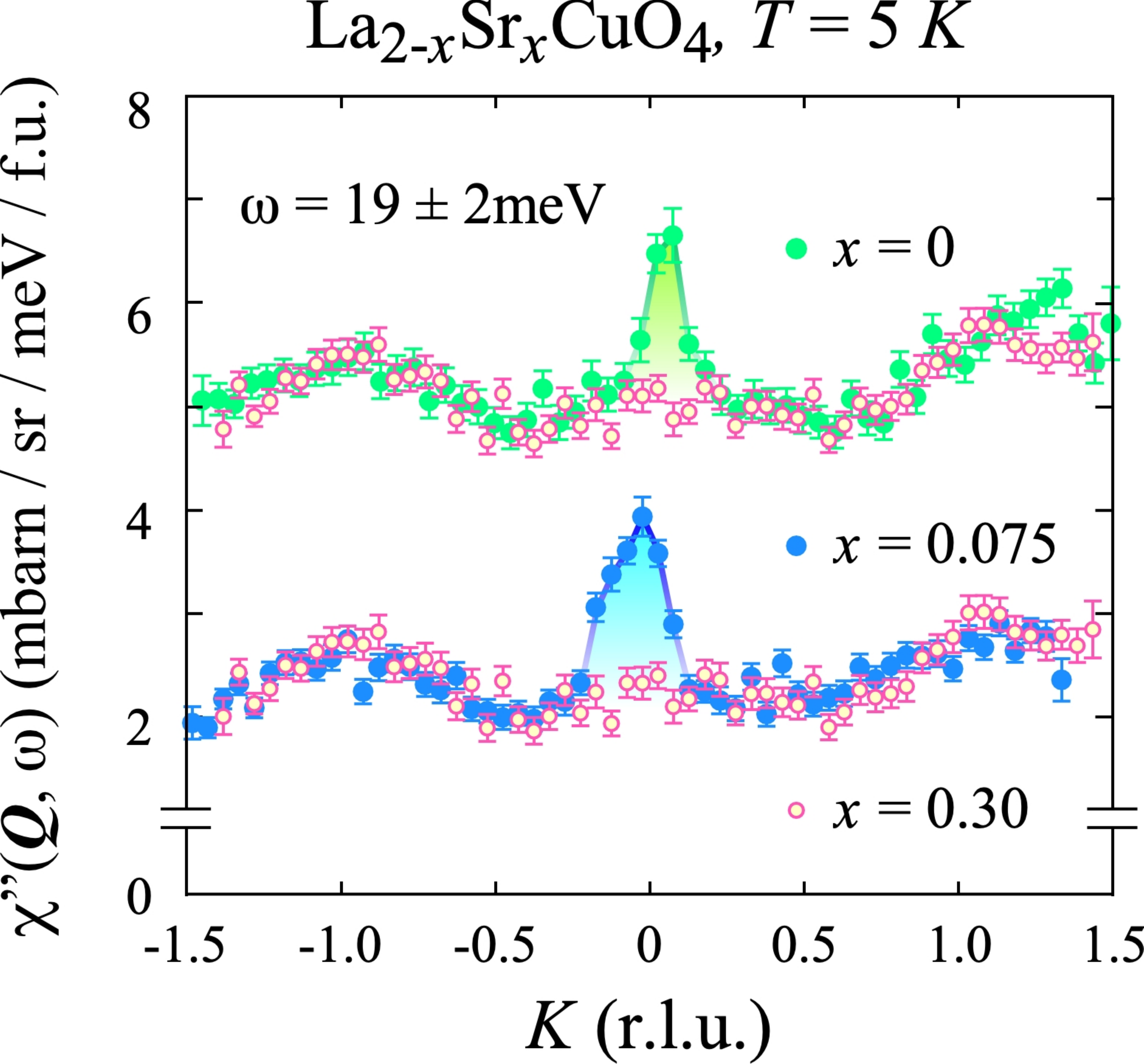}
\caption{
(Color online) The dynamical susceptibility of La$_{2-x}$Sr$_x$CuO$_4$ with $x$ = 0, 0.075, and 0.30 sliced at 19 meV with a width of $\pm$2 meV. The result for $x$ = 0 and overlaid values of $x$ = 0.30 are shifted upward by 3. The intensity was obtained by integrating over the ranges of $0.8 < H < 1.2$ and $2.5 < L < 3.5$. 
}
\label{fig2}
\end{figure}

\begin{figure}[t]
\includegraphics[scale=0.12]{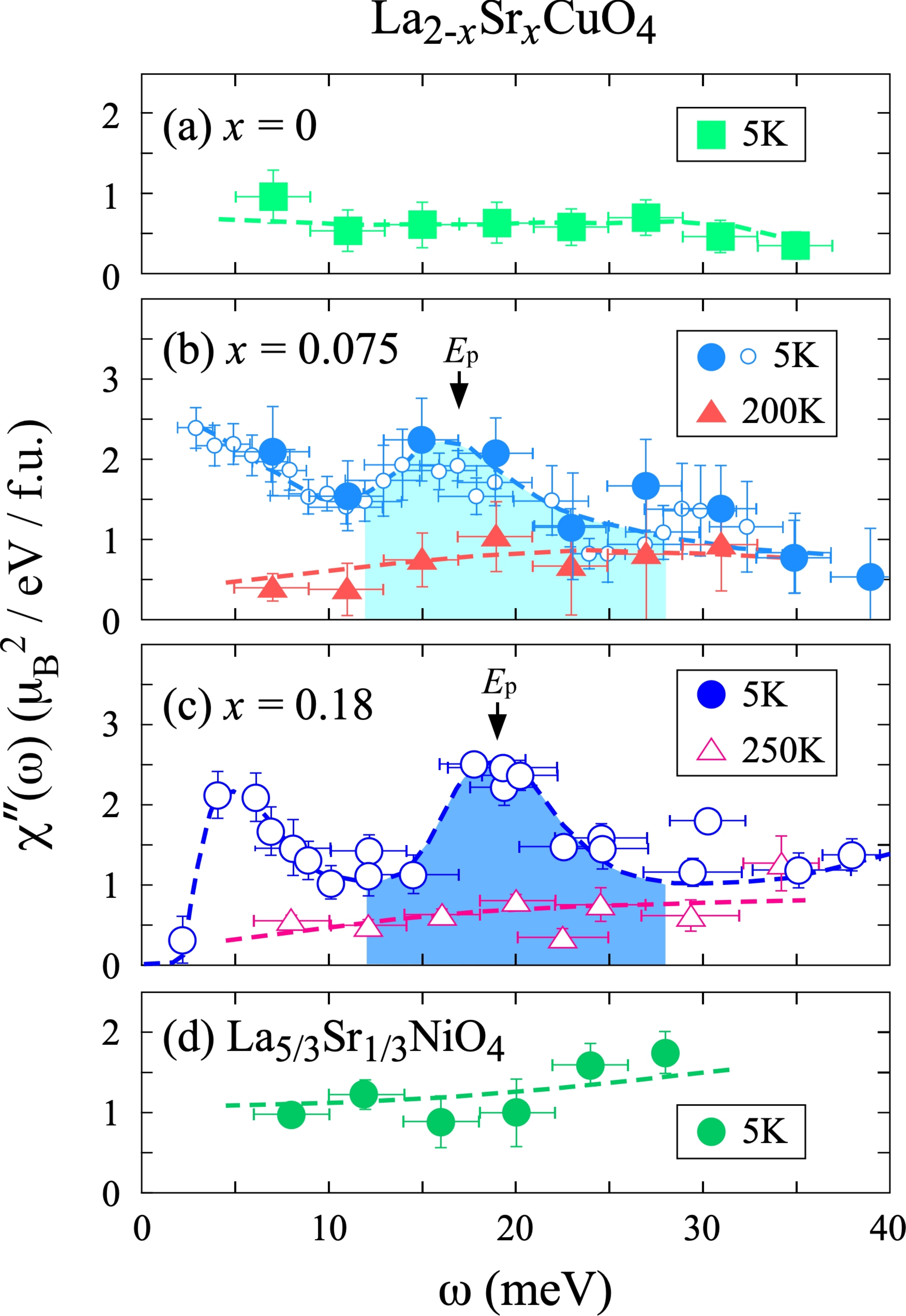}
\caption{
(Color online) Energy-dependencies of the local spin susceptibility $\chi ''(\omega)$ of La$_{2-x}$Sr$_x$CuO$_4$ with: (a) $x$ = 0, (b) $x$ = 0.075, (c) $x$ = 0.18, and (d) La$_{2-x}$Sr$_x$NiO$_4$ with $x$ = 1/3. Closed and open symbols in each figure represent the results obtained at fixed and unfixed $L$ values, respectively. $E_{\rm p}$ represents the peak position. The colored hatch is the area producing the energy integral $\chi ''(\omega)$ at low temperatures, as shown in Fig. \ref{fig4}. Dotted lines are guides to the eye. 
}
\label{fig3}
\end{figure}

In Fig. \ref{fig2}, the constant $\omega$ spectra at 19 meV are presented for LCO, LSCO(7.5), and LSCO(30). The results for LCO and the overlaid values of LSCO(30) are shifted upward by 3. The phonon intensities are in good alignment with each other. According to the INS studies reporting the intensity enhancement in the magnetic signal~\cite{Matsuura2012}, we first evaluate the bare magnetic intensity by subtracting the $\chi''({\bm Q}, \omega)$ of LSCO(30) from that of LCO and LSCO(7.5). Subsequently, we sliced the remnant magnetic spectra after the subtracting procedure at constant $\omega$ and fit the spectra using a Gaussian function to evaluate the \textit{\textbf{Q}}-integrated intensity ($\chi''(\omega)$). Details of this analysis are reported in Ref.~\onlinecite{wakimoto_proc}. 
The evaluated $\chi''(\omega)$ for LCO and LSCO(7.5) are shown in Figs. \ref{fig3}(a) and (b), respectively. As seen in Fig. \ref{fig3}(b), we clarify the presence of maximum $\chi''(\omega)$ in LSCO(7.5) near 16 meV. We note that phonon spectra for orthorhombic $x$ = 0 and 0.075 and those for tetragonal $x$ = 0.30 may differ from one another. However, the tilting mode of the CuO$_6$ octahedron, which is strongly connected with the crystal structure, is limited below $\sim$5 meV~\cite{wakimoto_phonon}. Therefore, the structural effects can be ignored when evaluating the bare magnetic signal above 10 meV, which is the focused energy range in the present study. 

Taking INS measurements while rotating the samples enables us to draw intensity maps in a wide $L$ range; thus the subtraction analysis at various $L$ from 0 to 7 r.l.u. could be carried out. 
As a result, we found that the bare magnetic spectra were independent from $L$ although the phonon intensity varied strongly with $L$~\cite{wakimoto_proc}. Therefore, we subsequently analyzed the data under the condition $k_i\,||\,c$ using a conventional method. In this measurement, the $L$ value was a function of $H$, $K$, and $\omega$. In Fig. \ref{fig3}(b), open circles represent $\chi''(\omega)$ with an unfixed $L$ value obtained after analytically subtracting the phonon intensity. The results agree with those evaluated at a fixed $L$, supporting the two-dimensional nature of $\chi''(\omega)$. Thus, the intensity enhancement is magnetic in origin. 

Based on these agreements, we analyzed the spectra measured with $k_i\,||\,c$ for slightly OD LSCO(18). 
As shown in Fig. \ref{fig3}(c), $\chi''(\omega)$ has an intensity maximum at $\sim$19 meV, and the peak-structure vanishes at 250 K. These results are quite similar to those for LSCO(7.5), even though $E_{\rm p}$ is higher than in LSCO(7.5). 
Therefore, the enhancement would be a characteristic feature in the SC phase at low temperatures. To further elucidate the origin of intensity enhancement, we examined LSNO(1/3), which has a diagonal stripe order.
As the intensity map at $L$ = 2 r.l.u. shows in Fig. \ref{fig1}(g), the spin excitations vertically stand at the incommensurate position with $K = \pm1/3$ r.l.u. and cross phonon branches lying horizontally at $\sim$14 and $\sim$20 meV. 
In Fig. \ref{fig3}(d), we present $\chi ''(\omega)$ for LSNO, which was evaluated with the same procedure as for LCO and LSCO(7.5). 
The enhancement of $\chi ''(\omega)$ was not seen at energies where optical phonons intersected. Combined with the presence of a peak-structure in $\chi ''(\omega)$ for SC LSCO, the absence of a similar structure in the doped LSNO(1/3) and undoped LCO suggests that the metallicity, in addition to spin and lattice dynamics, is a fundamental factor for the intensity enhancement.

\begin{figure}[t]
\begin{center}
	\includegraphics[scale=0.115]{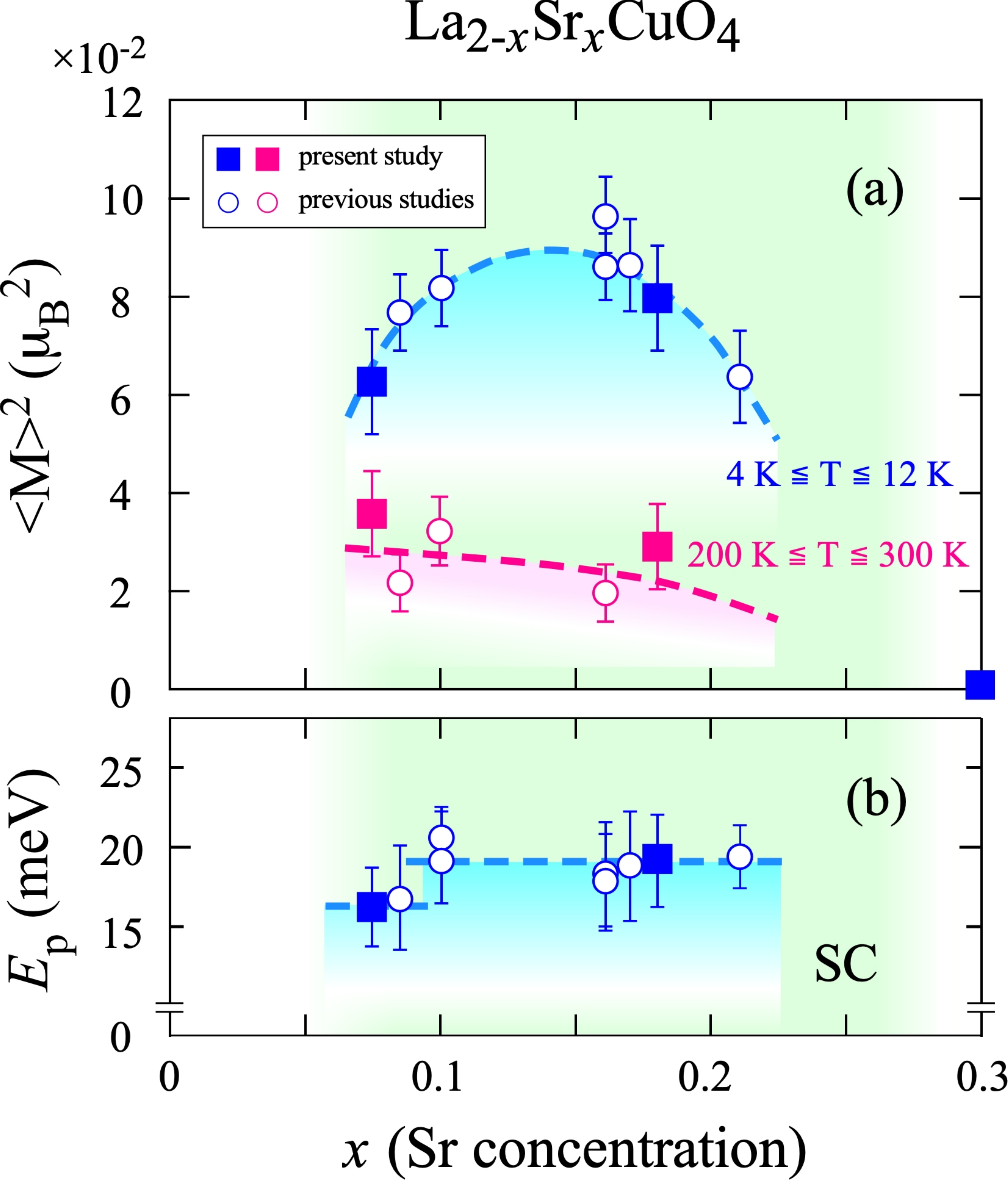}
\caption{(Color online) $x$-dependence of: (a) energy integrated local spin susceptibility from 12 to 28 meV at low (blue symbols) and high (pink symbols) temperatures, and (b) peak positions in the energy-dependence of local spin susceptibility for La$_{2-x}$Sr$_x$CuO$_4$. The values for the open symbols were evaluated from previously reported results~\cite{vignolle_HG, lipscombe_OD, lipscombe_localchi, Sato2020, Li2018}. The green area represents the superconducting phase. Dotted lines are guides for the eye.
}
\label{fig4}
\end{center}
\end{figure}

Figure \ref{fig4}(a) summarizes the mean squared moment ($\textless M \textgreater^2$) obtained by integrating $\chi ''(\omega)$ over the $\omega$-range between 12 and 28 meV, covering the broad peak in $\omega$-dependence of $\chi ''(\omega)$. We evaluated the values for LSCO with several doping levels from the reported $\chi ''(\omega)$ in the absolute unit~\cite{vignolle_HG, lipscombe_OD, lipscombe_localchi, Sato2020, Li2018}. All the samples except for $x$ = 0.075 were measured under the condition of $k_i\,||\,c$ and individual $E_{\rm i}$'s, and thus, the $L$ values at $E_{\rm p}$ were different from one another. 
However, we found a negligible effect of the $L$-component on $\chi ''(\omega)$; thus, this $x$-dependence of integrated value yields important information on the origin of the intensity enhancement. The $\textless M \textgreater^2$ at low temperatures shows dome-like $x$-dependence, although the data around 1/8-doping is missing. The similar $x$-dependence between $\textless M \textgreater^2$ and SC transition temperature ($T_{\rm c}$) in the broad SC phase suggests the interplay between intensity enhancement and superconductivity~\cite{Takagi1989}. Fig. \ref{fig4}(b) depicts $E_{\rm p}$ as a function of $x$. The $E_{\rm p}$ increases from $\sim$16 to $\sim$19 meV at $x$ $\sim$ 0.09 upon doping.

\begin{figure}
	\includegraphics[scale=0.118]{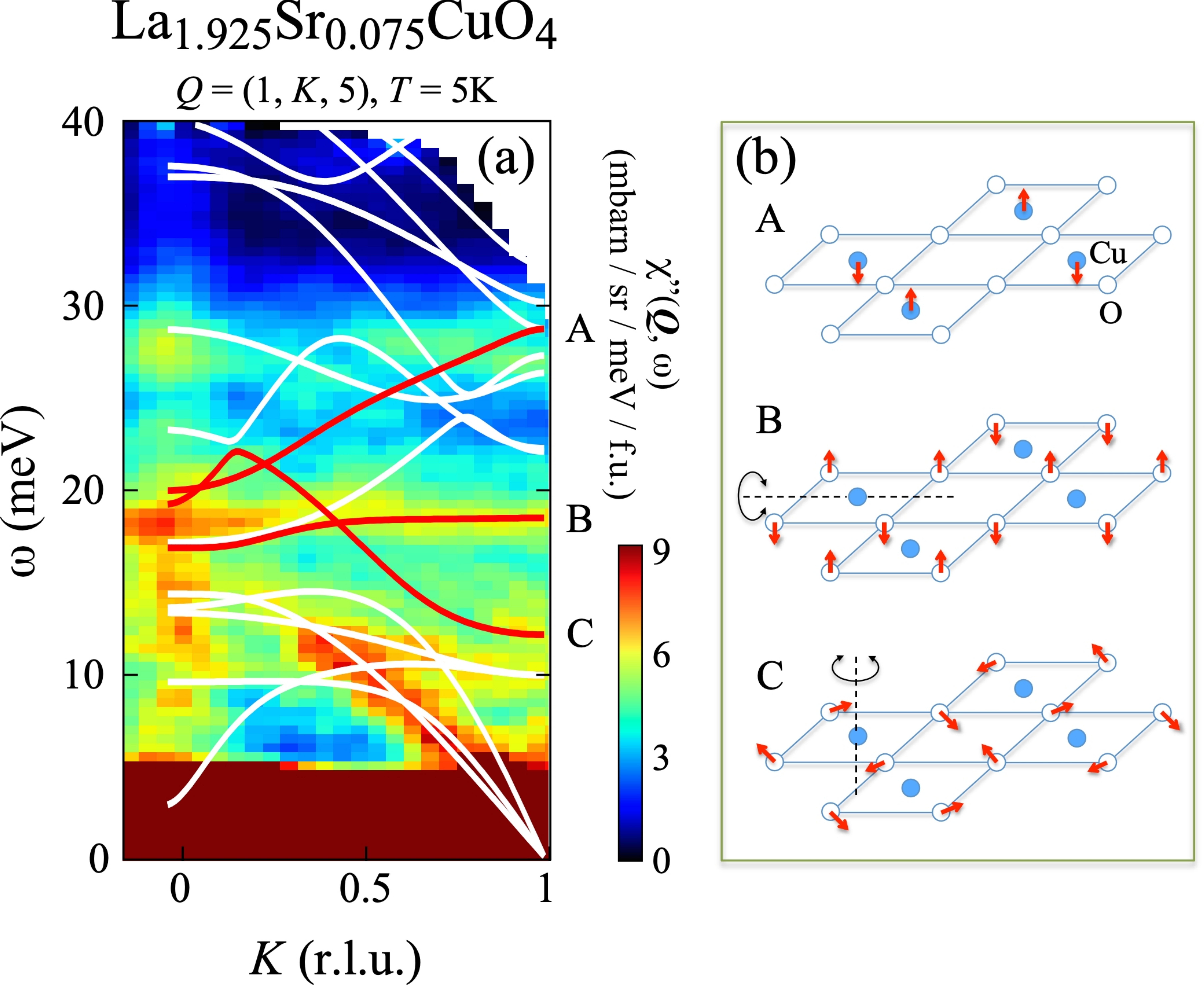}
\caption{
(Color online) (a) Intensity map of LSCO with $x = 0.075$ at 5~K in the $L=5$ zone, together with phonon branches calculated by the shell model~\cite{openphonon, phonon_param}.
The intensity map was obtained by integrating $H$ and $L$ from 0.8 to 1.2 and 4.5 to 5.5 r.l.u., respectively. 
(b) The displacement patterns of the CuO$_2$ plane at (1, 0) for the three highlighted phonon modes, labeled A, B, and C in (a).
}
\label{fig5}
\end{figure}

\section{Discussion}
In this study, our INS measurements on several characteristic samples provided new insights into the intensity enhancement. First, no intensity enhancement of phonons was observed for the OD LSCO(30) where low-energy IC magnetic fluctuations were absent\cite{wakimoto_OD1}. This observation implies that phonons cannot solely induce the emergent phenomenon and supports the importance of spin-phonon coupling for intensity enhancement\cite{Wagman2016}. Next, we confirmed two more important points; (1) the presence of a peak-structure of $\chi ''(\omega)$ with the maximum at 16 and 19 meV in UD and slightly OD LSCO, respectively, and (2) the absence of a peak-structure in the insulating systems having either commensurate (LCO）or incommensurate (LSNO(1/3)) low-energy spin excitations. By summarizing the present and previously reported $\chi ''(\omega)$ in the absolute value, the dome-shaped $x$-dependence of the integrated intensity around the peak energies was clarified. These results indicate that the peak intensity increases when increasing the itineracy of carriers and decreases in the OD region, where the magnetic correlation becomes weak~\cite{wakimoto_OD1}. Therefore, the intensity enhancement reflects the coupling of spin fluctuations with both phonons and carrier degree of freedom. The entanglement with carrier mobility supports the relationship between superconductivity and composite excitations of these degrees of freedom. 

One possible scenario for intensity enhancement through the interplay among spin, charge, and lattice dynamics is as follows. Deformation of the electronic band structure near the Fermi level occurs through electron-phonon interaction. Moreover, the resulting changes in nesting conditions cause the intensity enhancement of magnetic signals. This scenario is valid for metallic phases but not for the insulating phase. Furthermore, according to the Fermi surface nesting model~\cite{Norman2007}, the peak-structure could vanish at high temperatures where the anisotropic gap structure near the Fermi surface (the nesting condition) degrades. The evidence of an itinerant nature of the low-energy IC spin excitations of SC LSCO was reported by a recent neutron scattering experiment~\cite{Sato2020}. This is consistent with our results, based on the present findings that spin dynamics correlate with itinerant holes. 

An x-ray scattering measurement reported the existence of dynamic short-ranged charge orders that couple with phonons in LBCO with $x$ = 1/8, even above the onset temperature of the static charge order~\cite{Miao2019}. Development of a slow charge and lattice fluctuations upon cooling was more recently revealed by nuclear magnetic resonance measurements on LBCO with $x$ = 1/8~\cite{Singer2020}. 
Assuming that the one-dimensional stripe alignment of doped holes is essential for the coupling between phonons and holes, we consider possible phonons lying around 19 meV. Calculated phonons for the parent LCO, based on the shell model~\cite{openphonon, phonon_param}, are shown in Fig. \ref{fig5} with the intensity map in the $L$ = 5 r.l.u. zone of LSCO(7.5). This $L$ value is selected for a clear comparison, where phonons have higher intensities. Among the phonons crossing the spin excitations around 19 meV, we focus on three modes labeled as A, B, and C in Fig. \ref{fig5}, which shows the significant motions of Cu and O atoms of the CuO$_2$ plane. The motions of atoms for each mode are illustrated in the right panel of Fig. \ref{fig5}. Phonon modes A and B are associated with the out-of-plane motion of Cu or O atoms, producing a potential for trapping the charge stripes along the Cu-O bond direction, and hence, could be coupled with charges. The presence of two $E_{\rm p}$'s, shown in Fig. \ref{fig4}(b), suggests that the two different phonons contribute to the intensity enhancement, which is consistent with the result of the calculation. The variation of the strength of the interplay between charge/spin dynamics and these phonons could depend on the doping level, resulting in an increase in $E_{\rm p}$ with doping.  We note that the experimentally evaluated $E_{\rm p}$ is slightly lower than the calculated energy of the two phonons at $K$ = 0. This difference is possibly due to the phonon softening caused by carriers in the doped system. 

Finally, we discuss the intensity enhancement in other cuprates. Angle-resolved photoemission spectroscopy (ARPES) measurements of cuprate superconductors revealed a kink structure in the electronic dispersion at 60 and 40 meV~\cite{lanzara, graf, zhou_LSCO_arpes} as a result of electron-phonon coupling. The kinks at 40 and 60 meV originated from the coupling of holes with the oxygen buckling and half-breathing modes of the CuO$_2$ plane, respectively~\cite{cuk_arpes}. High-resolution ARPES measurements on the SC phase of Bi$_2$Sr$_2$CaCu$_2$O$_{8+\delta}$ revealed a new kink structure in the electronic dispersion at $\sim$16 meV and a decrease in anomaly strength upon doping ~\cite{anzai}. 
These results support the existence of low-energy phonon modes that couple with carriers and the doping evolution of the low-energy IC spin excitations. 
Meanwhile, the intensity enhancement of the magnetic signal was not reported for OP YBa$_2$Cu$_3$O$_{6+\delta}$~\cite{Hinkov2007} and OP HgBa$_2$CuO$_{4+\delta}$~\cite{Chan2016a, Chan2016b} using INS measurements, possibly due to the opening of a spin gap with large gap energy. That is, although optical phonons exist, the intensity enhancement cannot occur due to the absence of a magnetic signal below the gap energy, similar to LSCO(30). Investigating the spin-phonon coupling in the electron-doped system is essential to understand the microscopic mechanism of the intensity enhancement and its relationship with superconductivity. 
INS measurement on SC Nd$_{2-x}$Ce$_x$CuO$_4$ and (Pr, La)$_{2-x}$Ce$_x$CuO$_4$, which exhibit low energy commensurate spin excitations~\cite{Yamada2003, Fujita2008} and charge-density-wave order~\cite{Neto2015, Neto2016}, could provide vital information to test  theoretical works.

\section{Summary}
To clarify the origin of the peak-structure in the local magnetic susceptibility $\chi''(\omega)$ at 16--19 meV of LSCO, we examined the spin excitations in LCO, LSCO(7.5), LSCO(18), LSCO(30), and LSNO(1/3). 
We confirmed the peak-structure in SC LSCO(7.5) at $\sim$16 meV and LSCO(18) at $\sim$19 meV for 5 K, where the spin excitations superimposed on optical phonon branches. No intensity enhancement of the phonon and no peak-structure in $\chi''(\omega)$ were detected for LSCO(30), where low-energy spin excitations were absent as well as for the insulating LCO and LSNO(1/3). 
Furthermore, the energy integrated $\chi''(\omega)$ covering the $E_{\rm p}$'s showed a dome-shaped $x$-dependence similar to the $x$-$T_{\rm c}$ relation in LSCO. Based on the shell model, we determined that the candidate phonons coupled with the stripe-aligned holes. These results suggest that the peak-structure is due to the interplay among spin, charge, and lattice dynamics.

\begin{acknowledgments}
This work was performed under the Inter-University Cooperative Research Program of the Institute for Materials Research, Tohoku University (Proposal Nos. 15K0108, 16K0036, 17K0088, 18K0076, and 19K0058), and supported by JSPS KAKENHI under Grant Nos. 25390132 and 16H02125. The experiments at 4SEASONS were carried out under Project Nos. 2012P0201, 2013P0201, and 2014P0201. 
\end{acknowledgments}

\end{document}